\documentclass[12pt]{iopart}
\usepackage{graphicx}
\usepackage{citesort}

\begin{document}
\title{High Pressure Measurements of the Resistivity of  $\beta$-YbAlB$_4$}

\author{
T~Tomita,
K~Kuga,
Y~Uwatoko
and S~Nakatsuji
}

\address{Institute for Solid State Physics,
The University of Tokyo,
Kashiwanoha, Kashiwa, Chiba 277-8581, Japan}

\ead{ttomita@issp.u-tokyo.ac.jp and satoru@issp.u-tokyo.ac.jp }

\begin{abstract}
The electric resistivity $\rho(T)$ under hydrostatic pressure up to 8 GPa was measured above 2 K using a high-quality single crystal of the Yb-based heavy fermion system $\beta$-YbAlB$_4$. We found pressure-induced magnetic ordering above the critical pressure $P_{\rm c} \approx $ 2.4 GPa. This phase transition temperature $T_M$ is enhanced with pressure and reaches 30 K at a pressure of 8 GPa, which is the highest transition temperature for the Yb-based heavy fermion compounds. In contrast, the resistivity is insensitive to pressure below $P_c$ and exhibits the $T$-linear behavior in the temperature range between 2 and 20 K. Our results indicate that quantum criticality for $\beta$-YbAlB$_4$ is also located near $P_{\rm c}$ in addition to the ambient pressure.

\end{abstract}

\section{Introduction}

Quantum-critical phenomena in the Yb-based compounds have not been well explored to date and thus have attracted great interest for possible novel quantum criticality produced by different electronic configurations between the electron-like 4$f^{1}$-Ce$^{3+}$ and hole-like 4$f^{13}$-Yb$^{3+}$.
As a prototype of quantum critical phenomena in the Yb-based materials, non-Fermi-liquid of YbRh$_2$Si$_2$ has been well studied \cite{Review3}.
This material exhibits field-tuned quantum criticality by suppressing the low N\'{e}el temperature $T_N$ (= 70 mK) by magnetic field.  
However, except the recently discovered $\beta$-YbAlB$_4$, the Yb-based HF superconductivity has never been observed, neither at ambient conditions nor under hydrostatic pressure.

The intermetallic compound $\beta$-YbAlB$_4$ is the first Yb-based HF superconductor with the transition temperature $T_c$ (= 80 mK) and exhibits quantum criticality at zero field \cite{Review6,Review7}.
Hence, $\beta$-YbAlB$_4$ is one of the best systems to study quantum critical phenomena at ambient pressure\cite{Matsumoto2011Science}.
For example, the temperature dependence of the zero-field resistivity exhibits non-Fermi-liquid behavior, i.e., $T$-linear dependence from 4 K to 0.8 K; $T^{3/2}$ dependence from 0.8 K down to $T_c$ \cite{Review6,Review7}.
Moreover, in the temperature range $T_c$ $<T<$ 2 K, the observation of divergent susceptibility $\chi_{c} \propto T^{-1/2}$ under a low field of 50 mT applied along the $c$-axis and $-\ln T$ dependence of the magnetic part of the specific heat $C_{\rm M}/T$ also strongly suggests that this material has unconventional QCP, where standard theory based on spin density wave type instability is inapplicable.
In addition, strong valence fluctuation was observed in $\beta$-YbAlB$_4$ (Yb$^{\sim2.75+}$), in comparison with other Yb-based QC materials such as YbCu$_{5-x}$Al$_x$ (Yb$^{\sim2.95+}$) and YbRh$_2$Si$_2$ (Yb$^{\sim2.9+}$) \cite{Review25,Yamamoto}.
Hence, the valence fluctuations possibly play an important role in understanding this unconventional quantum criticality in $\beta$-YbAlB$_4$ and thus exotic behavior may appear through controlling parameters such as pressure and chemical doping.

A study of pressure as one of the control parameters may reveal the phase diagram near QCP in both Ce-based and Yb-based HF systems.
In the Yb-based compounds, generally, 4$f$ moments are known to become more localized with application of pressure.
Hence, with increasing pressure, a long-range magnetic order is expected to be stabilized as observed in YbRh$_2$Si$_2$ and YbCo$_2$Zn$_{20}$ \cite{Review10,Review11}, in sharp contrast with their Ce-based counterparts. In the case of $\beta$-YbAlB$_4$, it is highly interesting to see how the unconventional zero-field quantum criticality observed at ambient pressure is associated with a magnetic order expected to emerge under high pressure.
We report here the observation of pressure-induced magnetic order using high-quality single crystals $\beta$-YbAlB$_4$ and discuss pressure-tuned quantum phase transition.

\section{\label{sec:level2}Experimental details}

Single crystals of $\beta$-YbAlB$_4$ were grown by the aluminum self-flux method \cite{Review12}. To obtain high-quality single crystals, several crystals were selected using the residual resistivity ratio RRR=$\rho_{ab}(\mathrm{300 K})/\rho_{ab}(\mathrm{0 K})$.
We spot-welded electrical contacts to the surface of crystals using 20 $\mu$m\textit{$\phi$} Au wires.
The temperature dependence of the electrical resistivity under various pressures up to 8 GPa was measured for the selected single crystals $\beta$-YbAlB$_4$ over RRR = 200 (residual resistivity $\sim$ 1 $\mu \Omega$ cm) in the temperature region between 2 and 300 K using a cubic-anvil-type pressure cell.
Hydrostatic pressure up to 8 GPa was applied using a pressure transmitting media, Daphne oil 7373 \cite{Review13}.

\section{\label{sec:level3}Results and discussion}

  \begin{figure}[h!]
\includegraphics[scale =1]{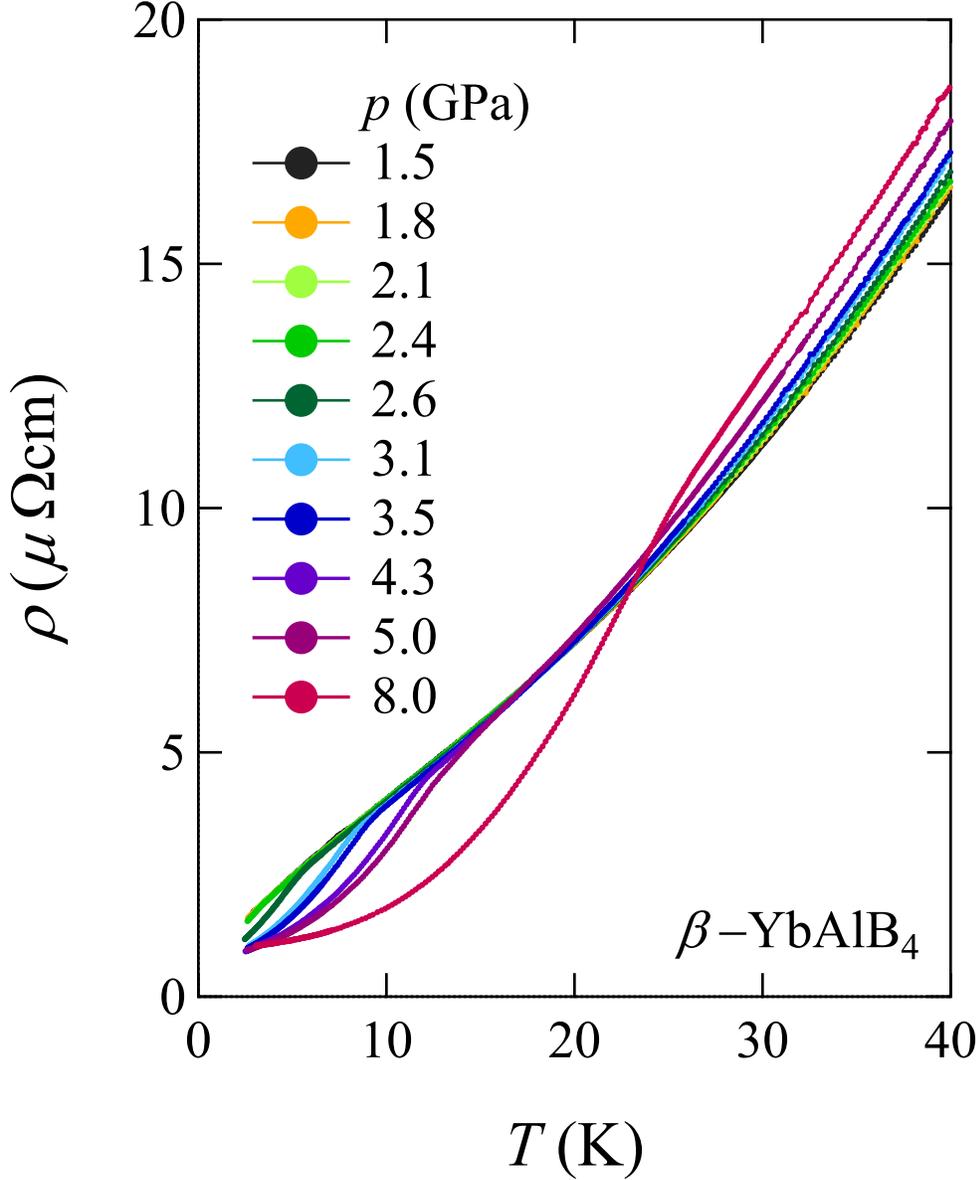}
\caption{ Pressure dependence of the in-plane resistivity $\rho_{ab}$ of $\beta$-YbAlB$_4$ (RRR=200) up to pressure of 8 GPa.
The inset shows the resistivity over a wide temperature range from 2 to 300 K under various pressures between 0 and 8 GPa. 
See text for details.}
  \label{fig-1}
  \end{figure}

Figure \ref{fig-1} displays the in-plane electrical resistivity $\rho_{ab}(T, p)$ of $\beta$-YbAlB$_4$ (RRR=200) in a pressure range of 0$\le p \le$ 8 GPa measured using the cubic anvil pressure cell with the pressure medium of Daphne 7373.
While the temperature dependence of the resistivity $\rho_{ab}(T)$ is almost independent of pressure up to 2.1 GPa, $\rho_{ab}(T)$ starts showing a kink above $P$= 2.4 GPa.
The kink may well arise from a magnetic phase transition as the gradual drop in the resistivity is normally associated with the loss of spin-disordered scattering due to magnetic ordering.
The kink temperature $T_M$ gradually increases with the application of pressure and reaches 30 K under 8 GPa.
The enhancement of $T_M$ with pressure is expected in a magnetically ordered Yb Kondo lattice compound, as we discussed above, and is in sharp contrast with the decrease in $T_M$ with pressure in Ce-based compounds.
The electrical resistivity at 300 K gradually increases with pressure, forming a maximum value at 6 GPa and then decreases at $P<$ 8 GPa.
The temperature slope change of the resistivity $\rho_{ab}(T)$ appears around 40 K corresponding to the peak found in the Hall coefficient $R_{\rm H}$, and may well come from the formation of the coherent state \cite{EoinPRL2012}. In particular, near this $T$ range close to 40 K, a systematic change in $\rho(T)$ as a function of pressure was observed.
The sharper increase of the kink temperature observed near $P_{\rm c} \approx 2.5$ GPa indicates that the magnetically ordered phase is not connected to the SC phase and is separated from the quantum criticality observed for $B$=$T$=0 at ambient pressure \cite{Matsumoto2011Science,Review15}.
Almost no change in resistivity is observed in the pressure rage between 0 and 2 GPa.
However, the magnetic transition of $\beta$-YbAlB$_4$ was suddenly found at $P>$2 GPa. Significantly, the phase transition of $\beta$-YbAlB$_4$ reaches 30 K under 8 GPa.  Such high magnetic transition temperatures over 10 K has never  been achieved in Yb-based HF materials, e.g.,
YbInCu$_4$ ($T_{\rm{M}}$=2.4 K at $p$=4 GPa \cite{Review22}) ,
YbCu$_2$Si$_2$ ($T_{\rm{M}}$=10 K at $p$=10 GPa \cite{Review29}),
YbRh$_2$Si$_2$ ($T_{\rm{M}}$=7 K at $p$=8 GPa \cite{Review4}), and
YbNi$_2$Ge$_2$ ($T_{\rm{M}}$=2 K at $p$=100 GPa \cite{Review30})
. Hence, the transition temperature of 30 K is the highest transition in the Yb-based HF compounds. In addition, the transition temperature is as high as that for CeRu$_2$Al$_{10}$ showing significantly enhanced $T_N$ in Ce-based HF compounds \cite{Muro}.

\section{\label{sec:level4}Conclusion}
A pressure-induced magnetic phase transition of $\beta$-YbAlB$_4$ above $P_{c} \approx $ 2 GPa has been observed by the electrical resistivity measurements.
In the intermediate pressure region 0$<P<$2 GPa, the resistivity is almost pressure independent and no magnetic order is found down to 2 K. Our resistivity measurements under pressure has revealed that non-Fermi-liquid state arises nearby the critical pressure of the magnetism, suggesting a magnetic quantum criticality near $P_{\rm c} = 2.4$ GPa. Furthermore, the dramatic difference in the phase diagram between those obtained by different pressure medium indicates that the magnetic quantum phase transition may be the first order. Given the fact that $\beta$-YbAlB$_4$ exhibits the zero-field quantum criticality under ambient pressure \cite{Review6,Review7}, 
several scenarios are possible.
(1) $\beta$-YbAlB$_4$ may have the 1st-ordered phase transition at $P_{\rm c}$ = 2.5 GPa such as a magnetic transition from a low moment magnetic state to a high moment magnetic state, as seen in the low pressure side of YbRh$_{2}$Si$_2$ near 2 GPa \cite{Review23,Review24} or a valence transition with changes in Yb valence as seen at the high-pressure side of YbRh$_{2}$Si$_2$ near 9 GPa \cite{Review10}.
(2) In the Ce-based HF superconductors, the superconducting phase and quantum criticality are connected to the AF magnetically ordered phase \cite{LonzarichNature}. In $\beta$-YbAlB$_4$, however, the SC phase and the ambient pressure quantum criticality may be separated from the magnetically ordered state. It is highly interesting to see if the non-Fermi liquid state at the ambient pressure forms a phase reaching to the critical pressure of the magnetism as has been observed in MnSi \cite{Review17}.
(3) $\beta$-YbAlB$_4$ may have two quantum critical regions at around ambient pressure and at around $P_{\rm c}$, which are separated by the Fermi liquid state.

In any case, a new type of phenomena in the Yb-based HF compounds is expected, and thus we need to further investigate the pressure dependence for $\beta$-YbAlB$_4$ at low temperatures to uncover the ground state evolution from the non-Fermi-liquid state and unconventional superconductivity to the magnetically ordered state under pressure.

\section*{Acknowledgements}
The authors wish to thank K. Matsubayashi for support in experiments and K. Ueda, K. Miyake and S. Watanabe for fruitful discussions. The authors are also grateful to N. Horie for experimental assistance.
This work has been supported by Grants-in-Aid (Grant No. No. 25707030 and 24740243),
and 
Scientific Research on Innovative Areas "Heavy Electrons" from MEXT of Japan (Grant No. 21102507).

\section*{References}
\bibliography{iopart-num}

\end{document}